\documentclass[final,5p,times]{elsarticle}
\usepackage{url}
\usepackage{amsmath}
\usepackage{bm}

\usepackage{booktabs}
\usepackage{tabularx}
\usepackage{algorithmicx}
\usepackage{algpseudocode}

\usepackage{amssymb,amsmath,algorithm}
\usepackage{microtype}
\usepackage{pifont}
\usepackage{subfigure}
\usepackage{multirow}
\usepackage{threeparttable}

\journal{Nuclear Physics B}

\begin{document}

\begin{frontmatter}



\title{Stateful protocol fuzzing with statemap-based reverse state selection}


\author[labelsysu]{Liu Yu}
\author[labelsysu]{Shen Yanlong}
\author[labelsysu]{Zhou Ying}
\affiliation[labelsysu]{organization={Sun Yat-Sen University, School of Electronics and Communication Engineering},
            city={Shenzhen},
            postcode={518107}, 
            state={Guangdong},
            country={China}}

\begin{abstract}
Stateful Coverage-Based Greybox Fuzzing (SCGF) is considered the state-of-the-art method for network protocol greybox fuzzing. During the protocol fuzzing process, SCGF constructs the state machine of the target protocol by identifying protocol states. Optimal states are selected for fuzzing using heuristic methods, along with corresponding seeds and mutation regions, to effectively conduct fuzz testing. Nevertheless, existing SCGF methodologies prioritise the selection of protocol states without considering the correspondence between program basic block coverage information and protocol states. To address this gap, this paper proposes a statemap-based reverse state selection method for SCGF. This approach prioritises the coverage information of fuzzy test seeds, and delves deeper into the correspondence between the basic block coverage information of the programme and the protocol state, with the objective of improving the bitmap coverage. The state map is employed to simplify the state machine representation method. Furthermore, the design of different types of states has enabled the optimisation of the method of constructing message sequences, the reduction in the length of message sequences further improve the efficiency of test case execution. By optimising the SCGF, we developed SMGFuzz and conducted experiments utilising Profuzzbench in order to assess the testing efficiency of SMGFuzz.The results indicate that compared to AFLNet, SMGFuzz achieved an average increase of 12.48\% in edges coverage, a 50.1\% increase in unique crashes and a 40.2\% increase in test case execution speed over a period of 24 hours.
\end{abstract}



\begin{keyword}


Stateful coverage-based greybox fuzzing
\sep Protocol fuzzing
\sep Greybox fuzzing

\end{keyword}

\end{frontmatter}


\section{Introduction}
\label{intro}
A network protocol is a set of rules for the exchange of network data, which specifies the format, response code and encryption method of data exchanged over a network. The implementation of network protocols forms the foundation of all network services, and their security is critical for all systems connected to them. Vulnerabilities in network protocols often lead to severe security issues. For example, the well-known Heartbleed vulnerability allowed attackers to read sensitive data from server memory, including user authentication credentials and private keys, resulting in the risk of user data leakage and identity theft. To enhance the security of network protocols, it is becoming increasingly important to conduct fuzzing test to uncover vulnerabilities in network protocols.

Fuzzing, since its proposal by Miller\cite{miller1990empirical} in 1990, has been favored by security researchers as an effective method for discovering vulnerabilities. Traditional fuzzing does not require testers to understand the specific internal workings of software, instead, it relies on continuously changing inputs to provoke unexpected behaviors in the software. As fuzzing techniques have evolved, simple black-box fuzzing has become insufficient to meet the needs of security researchers. Coverage-guided greybox fuzzing, pioneered by AFL\cite{afl} (American Fuzzy Lop), has gradually become an effective method for security researchers to discover vulnerabilities.

Although fuzz testing has achieved considerable success in identifying software vulnerabilities, there are still significant challenges in fuzz testing protocols\cite{bohme2020fuzzing}. Firstly, as the fuzz testing target, protocol implementations accept structured inputs. Secondly, fuzz testing inputs that do not conform to the protocol structure will be discarded by the testing target, rendering some test cases generated based on mutation methods ineffective. Secondly, protocol programs comprise numerous internal states, necessitating the generation of input sequences that can adapt to the protocol's state machine. In contrast to stateless testing targets, fuzz testing requires optimisation for the protocol's state machine. Finally, the protocol processing process necessitates network interaction, which significantly reduces the execution speed of the testing target. Consequently, the fuzzing test of protocol programs is inherently more challenging.

Currently, methods for fuzz testing protocols mainly include template-based black-box fuzzing and SCGF (stateful coverage-based greybox fuzzing). 

\textbf{Template-based black-box fuzzing.} Template-based black-box fuzzers require researchers to write protocol description templates corresponding to the protocol based on protocol specification documents such as Request for Comments (RFCs). Then, the fuzzer generates test cases according to the protocol description templates. Commonly used template-based black-box protocol fuzzing tools include Peach\cite{peach}, Boofuzz\cite{boofuzz}, and Kittyfuzzer\cite{kitty}. These tools construct protocol templates using protocol description primitives, which define the message format and state machine description of the protocol. The fuzzer generates test cases that comply with the protocol requirements based on these templates to conduct fuzz testing on the protocol under test. This method necessitates a considerable investment of time by security researchers in the construction of protocol description templates. As the complexity of protocol state machines increases, the difficulty of template development also increases, making it challenging to achieve large-scale automation testing.

\textbf{Stateful coverage-based greybox fuzzing.} SCGF is inspired by the CGF fuzzing method and guides protocol fuzz testing by constructing the state machine of the System Under Test (SUT). AFLNet\cite{pham2020aflnet} is the first fuzz tester based on SCGF. It identifies and records the states of the SUT through its response information, constructs the state machine of the SUT, guides the selection of test cases and mutation strategies based on the state machine, and finally mutates seeds to generate test cases. AFLNet has designed FAVOR, RANDOM, and ROUND-ROBIN state selection algorithms to choose an optimal state from those already fuzzed as the next target state for fuzzing. It then selects seeds based on the chosen state to achieve heuristic fuzzing based on state.

Subsequent research on SGFuzz\cite{ba2022stateful},  SNPSfuzzer\cite{li2022snpsfuzzer}, 
\\Snapfuzz\cite{andronidis2022snapfuzz}, Stateafl\cite{natella2022stateafl}, etc. inherited the state-guided algorithms from AFLNet\cite{pham2020aflnet} and made contributions in SUT state identification techniques, testing speed improvement, and other aspects. AFLNetLEGION\cite{liu2022state} further researched state-guided algorithms by extending the LEGION\cite{liu2020legion} algorithm to the state selection in protocol fuzz testing. It introduced more heuristic variables to further improve coverage. However, this more complex heuristic algorithm did not significantly improve the efficiency of the state selection algorithm. In AFLNetLEGION, the authors summarized the main reasons for this as follows: 1) the state identification model is not fine-grained enough; 2) the throughput is too low to unleash the full performance of the state selection algorithm.

Although the SCGF algorithm has achieved good results in guiding CGF to conduct fuzz testing on protocols, further enhancing the efficiency of the SCGF algorithm still faces two challenges:

\begin{enumerate}
    \item \textbf{State explosion in SUT}: 
    In existing SCGF approaches, researchers typically attempt to determine the program's state through return information or program execution information and construct the State Machine of the SUT  to guide fuzz testing. However, for complex programs, reconstructing the SUT's state space may become very difficult. Guiding protocol fuzz testing with a program's state machine is similar to guiding fuzz testing of binary programs with the program's execution flow graph. When the state space (or execution flow graph) of the testing target becomes highly complex, there is a possibility of state explosion, posing challenges for state selection.
    \item \textbf{Utilization of coverage information}: 
    In CGF, coverage information of the program is crucial. For example, AFL\cite{afl} culls the seeds in each testing round based on coverage information and scores seeds further based on coverage and execution speed factors to select optimal seeds for testing. In existing SCGF algorithms, the fuzz tester first constructs the state machine of the SUT based on response information (AFLNet\cite{pham2020aflnet}, SNPSfuzzer\cite{li2022snpsfuzzer}) or program execution information (Stateafl\cite{natella2022stateafl}, SGFuzz\cite{ba2022stateful}), selects the target states for fuzzing from the state machine, and then proceeds with seed selection. This state selection method is deficient in that it fails to utilise the potential logical relationship between basic block coverage information and program states. Consequently, paths that could lead to bitmap growth are not further explored by the fuzzer, and some seeds are not fully utilized.
\end{enumerate}

In this paper, we propose SMGFuzz, a SCGF fuzzer with statemap-based reverse state selection. It adopts a reverse state selection method for SCGF. SMGFuzz generates a state map by tracing the directed edges in the SUT state machine, in a manner analogous to that employed by AFL in the use of bitmaps. This approach enables the representation of intricate protocol state machines and addresses the issue of state explosion. During the fuzzing process, SMGFuzz prioritizes the coverage status of the bitmap for seed selection, based on which it further selects the target state that needs to be mutated. This process is contrary to the traditional SCGF approach, where states are selected first followed by seed selection. Hence, we term it as a \textit{reverse selection algorithm}. The reverse selection method enables SMGFuzz to thoroughly explore paths that contribute to bitmap growth, thereby achieving better seed utilisation rates than traditional SCGF. This enables SMGFuzz to achieve a more comprehensive coverage of bitmaps. Furthermore, SMGFuzz enhances the efficacy of the selection algorithm by refining the categorisation of states in Statemap.

In summary, the main contributions of this paper are the followings:

\begin{enumerate}[\textbullet]
    \item We analyzed the shortcomings of existing state-guided greybox fuzzing state selection algorithms and proposed a new representation method for the State Machine of the SUT based on Statemap.
    \item We designed a reverse state selection algorithm for protocol fuzz testing based on the Statemap. Leveraging the finer categorisation of protocol states, we devised a more efficient protocol fuzz testing state selection algorithm.
    \item We implemented SMGFuzz and conducted comparative experiments with the original state selection algorithm of AFLNet\cite{pham2020aflnet}. We also utilized Snapfuzz\cite{andronidis2022snapfuzz} to accelerate both SMGFuzz and AFLNet, further evaluating the effectiveness of the reverse algorithm.
\end{enumerate}


\section{BACKGROUND}
\label{background}

\subsection{Coverage-based greybox fuzzing}

Coverage-based greybox fuzzing (CGF) is considered state-of-the-art in fuzzing technology, with fuzzing tools like AFL\cite{afl}, libFuzzer\cite{serebryany2016continuous}, and honggfuzzer\cite{honggfuzz} built upon CGF serving as efficient binary software fuzzing tools. The core idea of CGF is to use lightweight instrumentation to collect coverage information for each execution of test cases on the SUT. Based on the coverage information of seeds, CGF selects seeds that can trigger new paths and maintains a subset of the seed set that covers the program execution paths triggered by the current seed set. This subset serves as the active seed set for fuzz testing.

Compared to traditional black-box fuzz testing, CGF utilizes coverage information from program execution to select better seeds, providing a directional approach to seed mutation in fuzzing. Additionally, CGF employs a genetic algorithm-based binary seed mutation algorithm by scoring seeds and continuously performing hovac and splicing operations on superior seeds. This integration of coverage information into the fuzzing process makes fuzzing more intelligent.

Compared to symbolic execution and taint analysis, CGF employs a more lightweight program analysis approach. It aggregates program coverage information into a bitmap, avoiding the extensive computational overhead associated with control flow analysis in programs. Moreover, it efficiently retains useful program execution information for fuzzing. The bitmap is composed of directed edges representing execution paths between basic blocks, where each point represents an execution path from basic block $a$ to basic block $b$ (excluding conflictions). This map simultaneously preserves certain basic block and program structure information. Through this simple representation of program coverage, tools like AFL\cite{afl} greatly enhance the efficiency of target program execution analysis, leading to higher fuzzing throughput.

Although CGF has been proven to greatly enhance fuzz testing efficiency, it faces challenges when applied to stateful SUT. This is because CGF lacks an understanding of program states, making it difficult to represent them directly. As a result, it is challenging to apply CGF directly to fuzz testing stateful SUT.

\subsection{SCGF}
To address the issue of perceiving the SUT state machine in fuzz testing of stateful SUTs, AFLNet\cite{pham2020aflnet} initially proposed SCGF. SCGF identifies the state of stateful SUTs, constructs an internal state machine of the SUT, and selects target states from this state machine to guide fuzzing of stateful SUTs. Subsequent tools like stateafl\cite{natella2022stateafl} SNPSfuzzer\cite{li2022snpsfuzzer}, and Snapfuzz\cite{andronidis2022snapfuzz} inherit the SCGF algorithm from AFLNet\cite{pham2020aflnet}. These tools aim to guide fuzz testing of stateful SUTs and improve the accuracy of state recognition and testing efficiency originally achieved by AFLNet\cite{pham2020aflnet}.

\begin{figure}[H]
    \centering
    \includegraphics[width=0.48\textwidth]{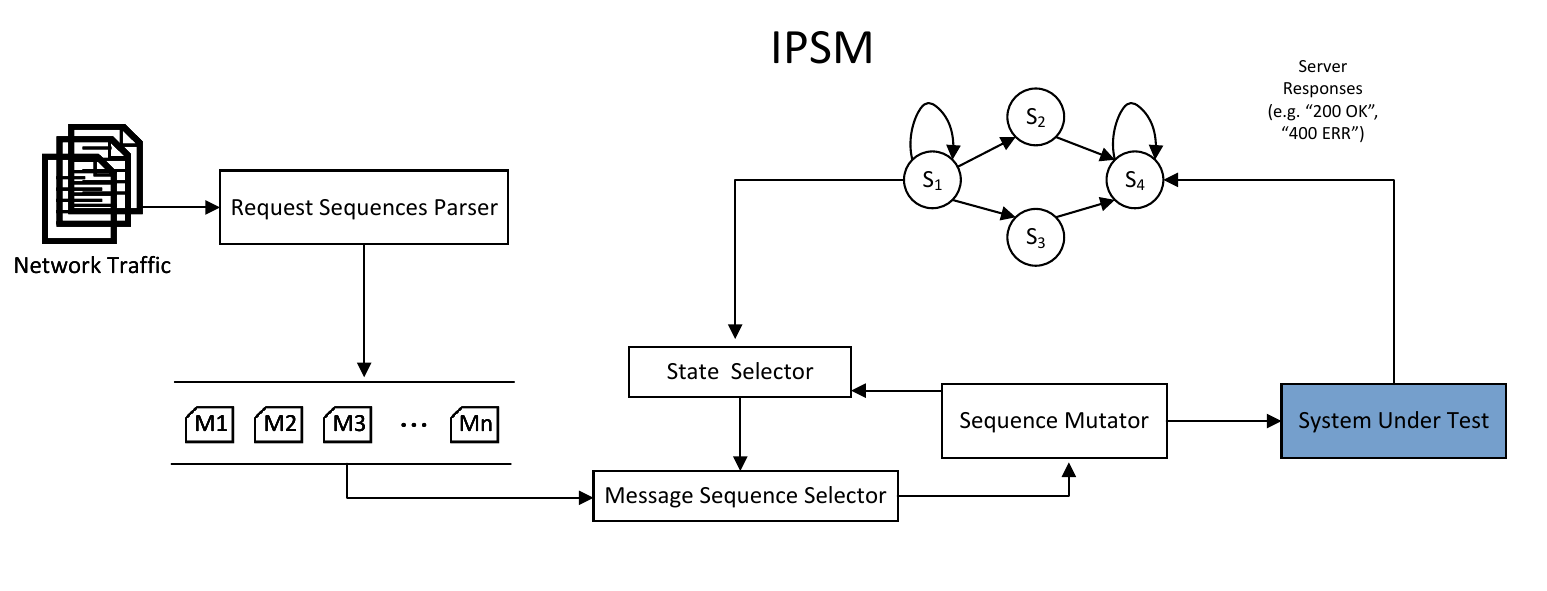}
    \caption{The architecture of AFLNet}
    \label{fig:aflnet}
\end{figure}

In SCGF, from a set of states \( S = \{s_1, s_2, s_3, \cdots\} \) and the corresponding message queue \( M = \{m_1, m_2, m_3, \cdots\} \), a state \( s \) is selected as the target state for fuzz testing in the current round. The message sequence \( M \) is mutated to \( M' \) in order to cover the selected state \( s \). To achieve this, SCGF divides the original sequence \( M \) into three parts:

1) Prefix \( M_1 \) needs to reach the selected state \( s \).

2) Candidate subsequence \( M_2 \) contains all messages that can be executed after \( M_1 \) but still remain in state \( s \).

3) Suffix \( M_3 \) is simply the remaining subsequence.

Thus, \( \langle M_1, M_2, M_3 \rangle = M \). The mutated message sequence \( M^\prime\) becomes \( \langle M_1, \text{mutate}(M_2), M_3 \rangle \). \( M_1 \) ensures that the mutated \( M' \) can cover state \( s \), while \( M_3 \) corresponds to the sequence for subsequent states. By mutating \( M_2 \), the fuzzer can explore new states that the SUT may trigger while in state \( s \). When the SUT accepts \( M' \) and triggers new program states, SCGF updates the state set \( S \) by adding the new states and records \( M' \) as an interesting seed for further testing.

In SCGF, the methods of selecting the target state \( s \) from \( S \) is divided into three types: FAVOR, RANDOM, and ROUND-ROBIN.

\begin{enumerate}[\textbullet]
    \item In the RANDOM method, SCGF randomly chooses the target state.
    \item In the ROUND-ROBIN method, SCGF iterates through the states in \( S \) one by one for testing.
    \item In the FAVOR method, SCGF first accumulates effective data of the SUT state machine through 5 rounds of ROUND-ROBIN. Then, it calculates the score of \( s \) heuristically and prioritizes selecting \( s \) with a higher score. It's noteworthy that the score of state \( s \) is inversely proportional to the number of executions (\#fuzz) and directly proportional to the number of states discovered (\#path). Once SCGF selects the target state \( s \), it then chooses a seed from the seed queue that can reach this state and performs mutation on it.
\end{enumerate}

AFLNetLEGION\cite{liu2022state} further refine the state selection algorithm of SCGF. It transforms the state machine graph obtained by SCGF into a tree-like representation, thereby distinguishing different paths to reach the state \( s \). AFLNetLEGION\cite{liu2022state} carrys out further optimizations on state selection and mutation strategies, resulting in improved coverage.

Currently, all SCGF algorithms rely on the construction of the SUT state machine, which is a prerequisite for the effectiveness of such algorithms. However, as protocol implementations become increasingly complex, building an accurate state machine for a complex protocol is a daunting task. Whether it is through protocol response codes (AFLNet\cite{pham2020aflnet}), program memory states (Stateafl\cite{natella2022stateafl}), or capturing program state spaces (SGFuzz\cite{ba2022stateful}), it is difficult to generate accurate program state machine graphs, and additional computational overhead is required to analyze the program state machine. Moreover, during the process of state selection in SCGF, bitmap information is not considered. The coverage information provided by the bitmap, representing the execution of corresponding seeds by the SUT, is obscured by the program state information, and there is no further exploration of the potential connection between the bitmap and program states.

\section{Motivation}
\subsection{Constructing a complete state machine for the SUT is always a challenge.}

\subsubsection{Time and resource consuming}
The SCGF method relies on a premise, which is the construction of the SUT state machine. Zhang et al.\cite{zhang2024survey} categorizes Communication Model Construction methods into Top-Down Approaches and Bottom-Up Approaches and analyzes the specific techniques used by various tools. Specifically, in tools that employ the SCGF method, AFLNet\cite{pham2020aflnet} utilizes protocol response codes as state markers. It continuously updates the current protocol's implemented protocol state machine (IPSM) using the response messages $R=\{r_1,r_2,r_3,\cdots\}$ obtained from the test sequence $M=\{m_1,m_2,m_3,\cdots\}$. This method has certain limitations: on one hand, different requests may receive the same response code (for example, in an HTTP-based protocol, successful GET and POST requests may both receive the same response code (200)); on the other hand, in some protocols, the response code may be unrelated to the current flow state (for example, in the HTTP2 protocol) \cite{{ba2022stateful}}. To address such issues, stateafl identifies the target program's state by tracking memory allocation and network I/O operations, and it marks the target state by constructing memory snapshots and applying fuzzy hashing. SGFuzz\cite{ba2022stateful} identifies and marks states using enumerated variables in the SUT.

In any case, acquiring the IPSM of the SUT requires a certain amount of computational resources and analysis of SUT-related information. These analyses require certain prerequisites, such as the target protocol's RFC document, or a sufficient amount of network traffic, or memory analysis of the target program. The consumption of computational resources and the limitations imposed by analysis conditions restrict the construction of the IPSM for existing SUTs, thereby also limiting the applicability of SCGF algorithms.

\subsubsection{Inprecision}
AFLNetLEGION\cite{liu2022state} is a more complex SCGF algorithm that relies on the construction of a tree-like exploration path based on the SUT's IPSM. It optimizes the effectiveness of state selection through path comparison. However, the distinction between paths depends on the accurate identification of preceding states. Any inaccuracies in state identification can lead to the appearance of incorrect paths or the failure to identify new paths.

\subsubsection{Discretizing the Abstraction of IPSM}
Using a complete state machine to guide SCGF is similar to using a control flow graph to guide fuzzing for stateless SUTs. An accurate state machine is crucial for fuzzing stateful SUTs. However, as the complexity of the target protocol and SUT increases, obtaining an accurate state machine becomes more challenging and resource-intensive. Inspired by greybox fuzzing, we can simplify the representation of the state machine while retaining essential state information, thereby simplifying the process of constructing and analyzing the SUT's state machine. Therefore, we have designed a statemap structure inspired by bitmaps for state analysis of stateful SUTs.

\subsection{Bitmap in SCGF}

\subsubsection{Relationship Between Bitmap and IPSM}
\begin{figure}[ht]
    \centering
    \includegraphics[width=0.48\textwidth]{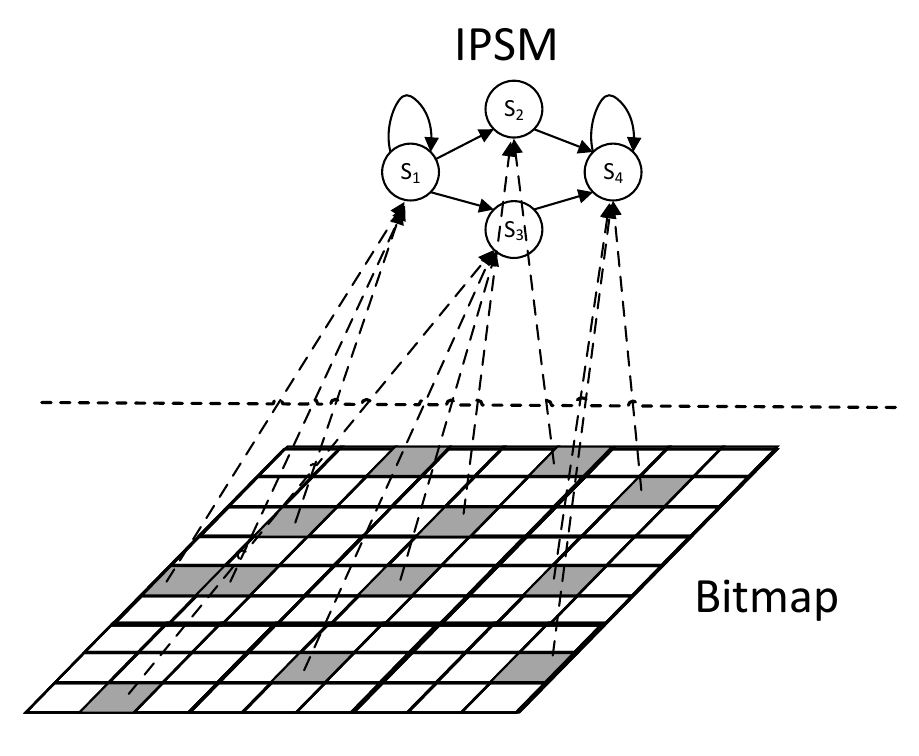}
    \caption{Caption of the image}
    \label{fig:ipsm}
\end{figure}
The bitmap in AFL\cite{afl} represents a mapping of hashed directed edges of program execution paths to memory space. Each byte in the bitmap represents the number of occurrences of a directed edge, assuming no collisions or overflows. Simplifying the bitmap to disregard repetition counts, each point in the \textit{bitmap\_mini} represents a directed edge appearing on the SUT execution path. The SUT's state machine serves as a high-dimensional abstraction of the program structure, indicating transitions between program states. Reverse-engineering the program into a CFG (control flow graph) establishes a mapping relationship between the state machine and the control flow graph.
Figure \ref{fig:ipsm} illustrates a mapping relationship between the bitmap and the IPSM, as both represent abstractions of the program structure.
In other words, the bitmap information implicitly contains program states. Similarly, triggering a certain program state $S$ will lead to the appearance of corresponding bits on the bitmap.
In SCGF, from a set of states \( S = \{s_1, s_2, s_3, \ldots\} \) and the corresponding message queue \( M = \{m_1, m_2, m_3, \ldots\} \), a state \( s \) is selected as the target state for fuzz testing in the current round.

\subsubsection{CGF as Markov Chain}
Marcel Böhme et al.\cite{bohme2016coverage} modeled CGF using a Markov Chain. In CGF, for a seed set \( T \), let \( S^+ \) denote the set of paths discovered from \( T \); \( S^- \subseteq S^+ \) represents the set of paths not discovered after fuzzing \( t \in T \). Then the state set \( S \) of the Markov chain is:

\begin{equation}
    S=S^{+} \cup S^{-}
\end{equation}
The state matrix \( P \) corresponding to the Markov Chain is defined as \( P=\left(p_{ij}\right) \). If path \( i \) is discovered by seed \( t_i\in T \), then \( p_{ij} \) represents the probability of seed \( t_i \) undergoing random mutation to generate path \( j \). 
\footnote{It is assumed that generating an input that exercises path \( j \) from the undiscovered seed \( t_i \) is as likely as generating from seed \( t_j \) an input that exercises the undiscovered path \( i \). Additionally, it is assumed that \( i \in S \) has no other undiscovered neighbors.}

For CGF, each round of testing seeds aims to attain global bitmap coverage for the current state to the fullest extent possible. Let \( S^b \) represent the subset of paths covering the bitmap for this round, and \( T^b \) denote the subset of seeds selected for \( S^b \) in this round. Then, the probability of triggering path \( j \) in this round for path \( j \) is:

\begin{equation}
    P_j^{\mathrm{C G F}}=\sum_{t_i \in T^b} p_{i j}
\end{equation}

\subsubsection{SCGF as Markov Chain}
In SCGF, let \( M^+ \) represent the set of protocol states discovered by the seed set \( T \), and \( M^- \nsubseteq M^+ \) denote the set of protocol states not discovered after fuzzing \( t\in T \). Then, the protocol state set corresponding to the Markov chain is:
\begin{equation}
    M=M^{+} \cup M^{-}
\end{equation}
Let \( D(t) \) denote the protocol target state reached after seed \( t \) is executed, where \( m\in M^+ \). \( T^{M^+} \) represents the set of seeds that cover \( M^+ \), i.e., \( M^+=\sum_{t\in T^{M^+}}{D(t)} \). 

For a round of testing in SCGF, the first step is to select the target state \( m\in M^+ \) with the corresponding seed set \( T^m\subseteq T \). After selecting the target state, SCGF will trim down the seed set by choosing seeds that cover edges in the bitmap and result in the protocol template state being \( m \), i.e., \( T^{mb}=T^b\bigcap T^m \). Then, a seed \( t\in T^{mb} \) is chosen for fuzzing. The probability of triggering path \( j \) in SCGF for this round can be expressed as:
\begin{equation}
    P_j^{\mathrm{S C G F}}=\sum_{t_i \in T^{m_n b}} p_{i j}\left(M^{+}=\sum m_n\right)
\end{equation}

For a round of testing in SCGF, under the condition that \( M^+=\sum m_n \), the selected seed set \( T^{Mb} \) is defined as: \( T^{Mb}=\left\{t_1\in T^b\bigcap T^{m_1}\right\}\bigcup\left\{t_2\in T^b\bigcap T^{m_2}\right\}\cdots\bigcup\left\{t_n\in T^b\bigcap T^{m_n}\right\} \). Thus, \( T^{Mb}\subseteq T^b \). For any target state \( m_n \), if there exist two seeds \( t_a \) and \( t_b \) that correspond to different covering subsets in the bitmap and satisfy \( D\left(t_a\right)=m_i \) and \( D\left(t_b\right)=m_i \), then \( T^{Mb}\subsetneq Tb \). Hence, if there exists a state with two corresponding seeds and the seeds have independent coverage areas in the bitmap, it is impossible to achieve full coverage of the bitmap in a single round of testing in SCGF.

\subsubsection{Reverse State Selection as Markov Chain}
In SCGF, coverage of the bitmap is prioritized, and reverse state selection is employed to enhance the utilization of bitmap information. For a seed \( t \), if it can cover two states \( m_i \) and \( m_j \), then during the execution of the target program, there exists a directed edge \( \langle a,b\rangle \) that realizes this state transition. Assuming no conflicts, there must exist a point \( \alpha \) in the bitmap corresponding to \( \langle a,b\rangle \).

In one testing round, \( T^b \) is selected as the seed set, and the corresponding protocol state coverage set is \( M^b \). Since all transitions between protocol states correspond to points in the bitmap, achieving coverage of the bitmap guarantees coverage of all protocol states, i.e., \( M^+ \subseteq M^b \). Based on this, a heuristic algorithm is used to select \( M^b \), involving the selection of target states for protocol mutation after selecting seeds, resulting in the final test seed set \( \mathcal{T}^b \) for this round.

For Reverse State Selection (RSS), the probability of triggering path \( j \) in one round of testing is:
\begin{equation}
    P_j^{\mathrm{R S S}}=\sum_{t_i \in \mathcal{T}^b} p_{i j}
\end{equation}

For a single round of testing, it is evident that \( P_j^{RSS} > P_j^{SCGF} \).

In RSS, state-based priority selection is abandoned in favor of prioritizing reverse state selection to ensure better coverage of bitmap-executed paths in each round, thereby increasing the likelihood of triggering undiscovered path \( j \) in a round. Additionally, heuristic selection of protocol target states is performed after seed selection, aiming for global optimization.

\subsection{Message Sequence Construction in SCGF}

\subsubsection{Increase in message sequence length }
In the existing SCGF, to ensure that the execution of the message sequence covers the target states and propagates the effects of mutation to subsequent states, SCGF needs to send all messages in the message sequence to the test target and complete one test cycle with each fuzzing execution. AFLNet\cite{pham2020aflnet} eliminates the seed trimming operation present in AFL. For SCGF, generating more unique state seeds will receive higher priority. Additionally, in non-deterministic mutation, AFL concatenates different seeds through continuous hovac and splicing, causing longer message sequences to receive higher execution priority in SCGF. Observations of seeds generated by AFLNet after running for 24 hours reveal that subsequently stored seeds may contain hundreds of messages, resulting in significantly decreased execution speed for such seeds.

\begin{figure*}[!h]
    \centering
    \includegraphics[width=\textwidth]{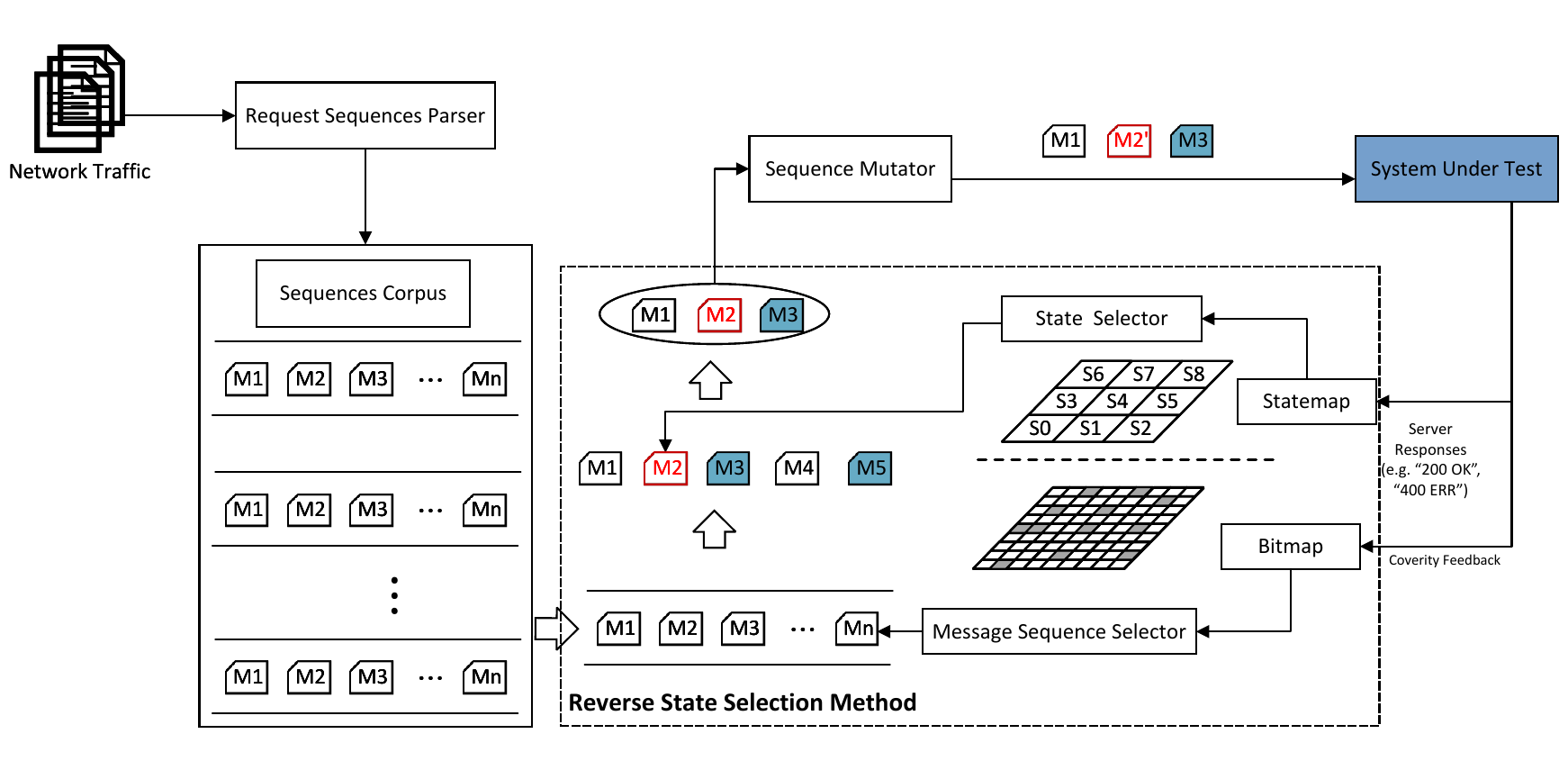}
    \caption{Architecture of SMGFuzz}
    \label{fig:overviewofsmgfuzzer}
\end{figure*}

\subsubsection{Improving Test Case Execution Speed}
AFLNet removes the trimming process present in AFL, ensuring that the message sequence will not be trimmed during the fuzzing process, without eliminating parts that do not contribute to coverage growth. The execution speed of seeds will decrease continuously during the fuzzing process. In the subsequently generated message sequence, there are a large number of duplicates (many repeated messages in one message sequence). For protocols, some messages can cause the TCP connection to be terminated. We can refer to this protocol state as the end state of the protocol. For example, in the FTP protocol, the quit command and incorrect login passwords will cause subsequent messages to need to re-establish the TCP connection. When fuzzing the protocol triggers the protocol end state, the effects of preceding messages on the SUT will not spread to subsequent states. In other words, it is meaningless to continue exploring subsequent states that trigger the end state for a single fuzzing. Therefore, we propose to define end states to divide the message sequence into independent subsequences, thereby reducing the execution of meaningless messages in single fuzzing, while ensuring the testing effectiveness and improving the average execution speed of test cases.

\section{Overview of SMGFuzz}

\subsection{Overall Design of SMGFuzz}
The overall architecture of SMGFuzz is illustrated in the Fig \ref{fig:overviewofsmgfuzzer}. Initially, SMGFuzz parses the network traffic to extract request sequences, which are then crafted into initial seeds. This process is similar to aflnet. As the fuzzing process begins, SMGFuzz constructs a statemap based on the response code after executing message sequences, adding state points to the statemap and marking end messages in the sequences. During each iteration of the fuzzing loop, SMGFuzz trims the Sequences corpus based on bitmap information, marks favored sequences, and selects the sequence to be used for the current fuzzing iteration. Then, based on the information from the statemap, SMGFuzz parses the message sequences and constructs the required sub-message sequences for the current fuzzing iteration based on the type of state point. Finally, SMGFuzz mutates the messages that need to be mutated for the current iteration according to the statemap and sends them to the SUT.

\subsection{Statemap Design}

\subsubsection{Statemap}

\begin{figure}[!h]
    \centering
    \includegraphics[width=0.48\textwidth]{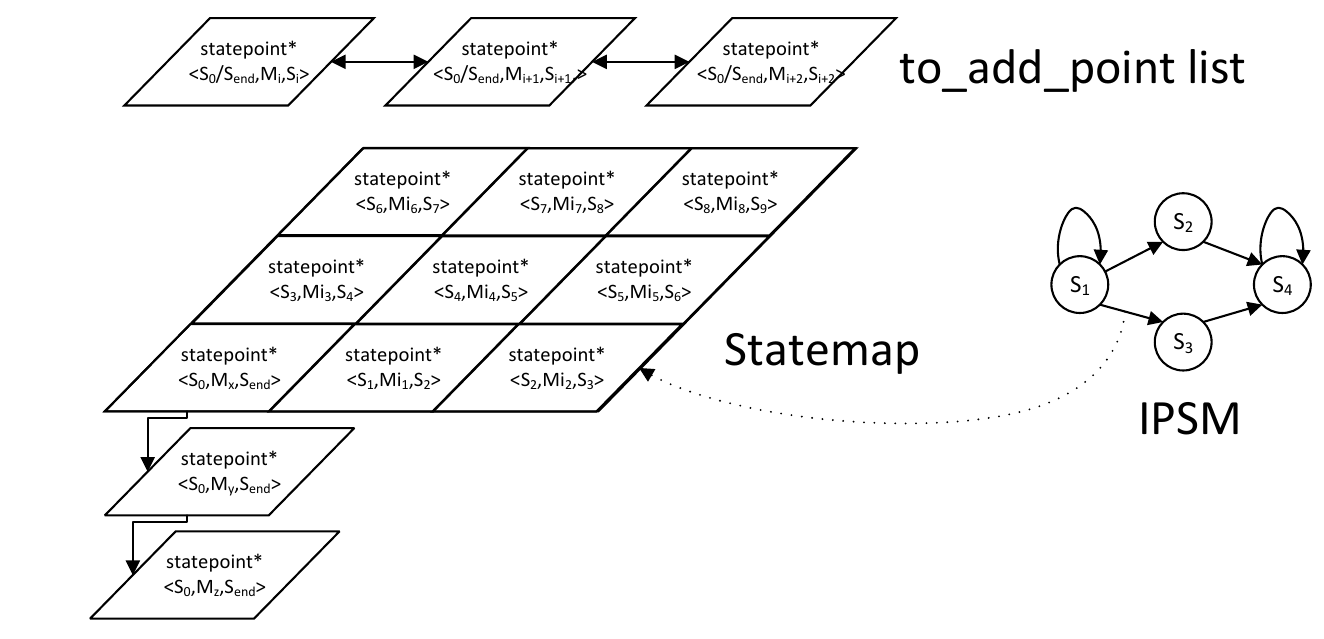}
    \caption{Statemap Design}
    \label{fig:statemap}
\end{figure}

The statemap consists of an array of pointers to statepoint objects, containing directed edges from the IPSM of the SUT. Each statepoint in the statemap is uniquely identified by the hash of the directed edge $ \langle S_n, S_{n+1} \rangle $.
Three types of \texttt{statepoints} exist: \texttt{POINT\_ZERO}, \texttt{POINT\_TO\_ADD}, and \texttt{POINT\_ADDED\_TO\_MAP}, as detailed in section \ref{statepoint}, explaining the differences between these three types of nodes. 
$S_n$ represents a state in the IPSM of the SUT, where $S_0$ and $S_{\text{end}}$ are two special states. 
The initial state of the SUT before receiving any messages is denoted as $S_0$, while $S_{\text{end}}$ represents the state where the SUT disconnects due to receiving abnormal messages or an end request (e.g., when an FTP server returns 530 for login errors).
\texttt{statemap[0]} is a special element in the statemap, pointing to a linked list of \texttt{POINT\_ZERO} type. All statepoints of this type are added to this linked list for subsequent use.

\subsubsection{Statepoint}
\label{statepoint}
\begin{figure}[t]
    \centering
    \includegraphics[width=0.48\textwidth]{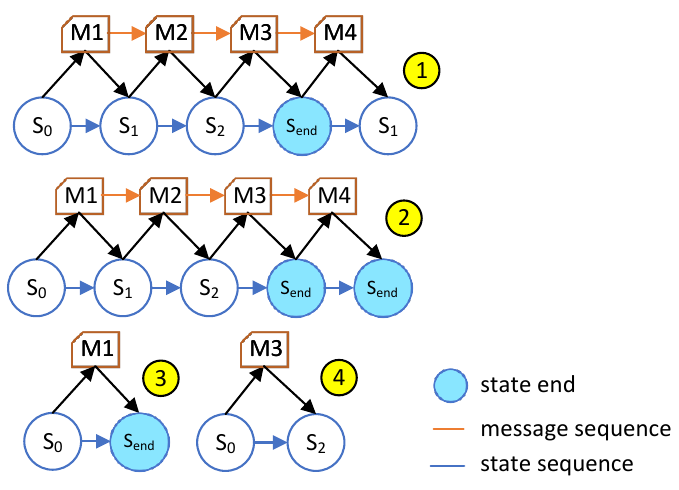}
    \caption{Statepoint Design}
    \label{fig:statepointdesign}
\end{figure}

A statepoint is an element in the statemap array, which stores all the information when the SUT transitions from one state $S_n$ to another state $S_{n+1}$. Here, $R_n$ represents the packet returned when transitioning to the $S_n$ state, $M_n$ represents the message sent in the $S_n$ state, and $R_{n+1}$ identifies the packet returned in the $S_{n+1}$ state after the transition. The statepoint also includes fuzz-related information, such as the type of the point, whether it has been fuzzed as a testing target, the number of times it has been fuzzed, the number of seeds generated at this point, and the set of all seeds that can trigger this point. This information will be used in subsequent statemap initialization and reverse selection algorithms.

Since statepoints represent the directed edges in the SUT's IPSM and introduce the initial state $S_{\text{0}}$ and the terminal state $S_{\text{end}}$, there are some special types of statepoints. Depending on the different initial and target states, we categorize statepoints into three types for later use in statemap initialization and statemap-based reverse selection algorithms.

\textbf{Definition:}
\begin{itemize}
    \item \textbf{Valid state transition}: A state transition where the source state is not $S_{\text{0}}$ or $S_{\text{end}}$.
    \item \textbf{Invalid state transition}: A state transition where the source state is $S_{\text{0}}$ or $S_{\text{end}}$.
\end{itemize}

\textbf{POINT\_ZERO}: Statepoints with an initial state of $S_{\text{0}}$ or $S_{\text{end}}$ and a target state of $S_{\text{end}}$, such as $\langle S_{\text{end}}, S_{\text{end}}\rangle$ and $\langle S_{\text{0}}, S_{\text{end}}\rangle$ as shown in Fig.\ref{fig:statepointdesign}(2) and  Fig.\ref{fig:statepointdesign}(3), respectively, often occur in login, authentication, and similar scenarios where the server rejects the connection due to incorrect usernames or passwords, resulting in an invalid state transition. 
These statepoints are stored in a linked list at the head of \texttt{statemap[0]} and are later used in conjunction with \texttt{POINT\_TO\_ADD} type nodes in the \texttt{to\_add\_point\_list} to conduct testing on the SUT.

\textbf{POINT\_TO\_ADD}: Statepoint with an initial state of $S_{\text{0}}$ or $S_{\text{end}}$ and no subsequent $M_{\text{n+1}}$ statepoint. Examples include  $\langle S_{\text{end}}, S_{\text{1}} \rangle$ as shown in  Fig.\ref{fig:statepointdesign}(1) and $\langle S_{\text{0}}, S_{\text{2}}\rangle $ as shown in Fig.\ref{fig:statepointdesign}(4), caused by the lack of subsequent messages leading to an invalid state transition. 
These statepoints are added to a to\_add\_point\_list doubly linked list, 
and upon forming valid state transitions with other nodes in the list with the same target state or \texttt{POINT\_ZERO} type nodes, they are removed from the to\_add\_point\_list and added to the statemap.

\textbf{POINT\_ADDED\_TO\_MAP}: Statepoint containing a valid state transition.

\subsubsection{Add statepoint to Statemap}
\label{addstatepoint}
In the fuzzing process based on CGF, seeds that trigger new paths will be added to the seed queue for subsequent use. In traditional SCGF, when seeds are added to the queue, the fuzz tester parses the message sequence and response sequence, adds the new states to the SUT's IPSM, and updates the IPSM. Unlike traditional SCGF, SMGFuzz treats state transitions as statepoints in the statemap and defines $S_{\text{0}}$ and $S_{\text{end}}$, thereby generating three different types of nodes. Therefore, algorithm ASTS is used to parse the message sequence and response sequence.

\begin{algorithm}[h]
    \label{alg:constructmessagesequence}
    \caption{Add Statepoint To Statemap (ASTS)}
    \begin{algorithmic}[1]
    \State \textbf{Input:}$M\!S$(messages sequence),$S$(state),$S\!C$(state count),
    \ $S\!M$(state map)
    \State \textbf{Output:} \ $S\!M$(state map)
    \State $M\!C$(message count) $\gets$ 0
    \State m $\gets$ NULL
    \State m\_prev $\gets$ NULL
    \If{$S\!C <=0$}
    \State \textbf{return}
    \Else
    \For{i from 0 to LEN($M\!S$)}
    \State m\_prev = m;
    \State $M\!C$++;
        \If{$M\!C$ $>$ state\_count}
        \State \textbf{return}
        \EndIf
        \If{m\_prev == NULL}
            \If{RESPONSE\_END(state[$M\!C$])}
            \State ADD\_PIONT\_TO\_ZERO\_LIST(state[$M\!C$-
            \Statex\qquad\qquad\qquad 1], state[$M\!C$])
            \State m = NULL;
            \Else
                \If{$M\!C$ = $S\!C$ -1}
                \State ADD\_PIONT\_TO\_ZERO\_LIST(state
                \Statex\qquad\qquad\qquad\quad\ [$M\!C$-1], state[$M\!C$])
                \EndIf
            \EndIf
        \Else
            \If{IN\_STATEMAP(state[$M\!C\!-1$],state[$M\!C$])}
            \State UPDATE($S\!M$);
            \Else
            \State ADD\_POINT\_TO\_STATEMAP(state[$M\!C$ - 
            \Statex\qquad\qquad\qquad 1], state[$M\!C$])
            \EndIf
        \EndIf
    \EndFor
    \EndIf 
    \end{algorithmic}
\end{algorithm}

Algorithm MSC 
takes the message sequence, state sequence, and the number of states as inputs. It traverses the message sequence sequentially, parsing statepoints with the m\_prev pointer pointing to the previous message and the m pointer pointing to the current message, and adds them to the statemap. Initially, both m\_prev and m are null, and then m is set to point to the header of the kl\_message. When m\_prev is null, the preceding state transition is either from $S_{\text{0}}$ or $S_{\text{end}}$. If the succeeding state is $S_{\text{end}}$, the state transition is of \texttt{POINT\_ZERO} type, and it is added to the state\_zero\_list; m\_prev is then set to null. If there is no succeeding message, the state transition is of \texttt{POINT\_TO\_ADD} type, and it is added to the to\_add\_point\_list. When m\_prev is not null, the point is added to the statepoint list maintained by the seed. The hash of $\langle S_{\text{src}}, S_{\text{dst}}\rangle$ is computed, and it is determined whether it already exists in the statemap. If the state transition already exists in the statemap, the current seed is added to the seed queue of that point; if not, $\langle S_{\text{src}}, S_{\text{dst}}\rangle$ is added to the statemap. Additionally, if $S_{\text{dst}}$ is $S_{\text{end}}$, m\_prev is set to null, and the point is recorded as a termination node in the state list maintained by the seed.

\subsection{Reverse State Selection}

\subsubsection{Construct Message Sequence}

\begin{figure}[h]
    \centering
    \includegraphics[width=0.48\textwidth]{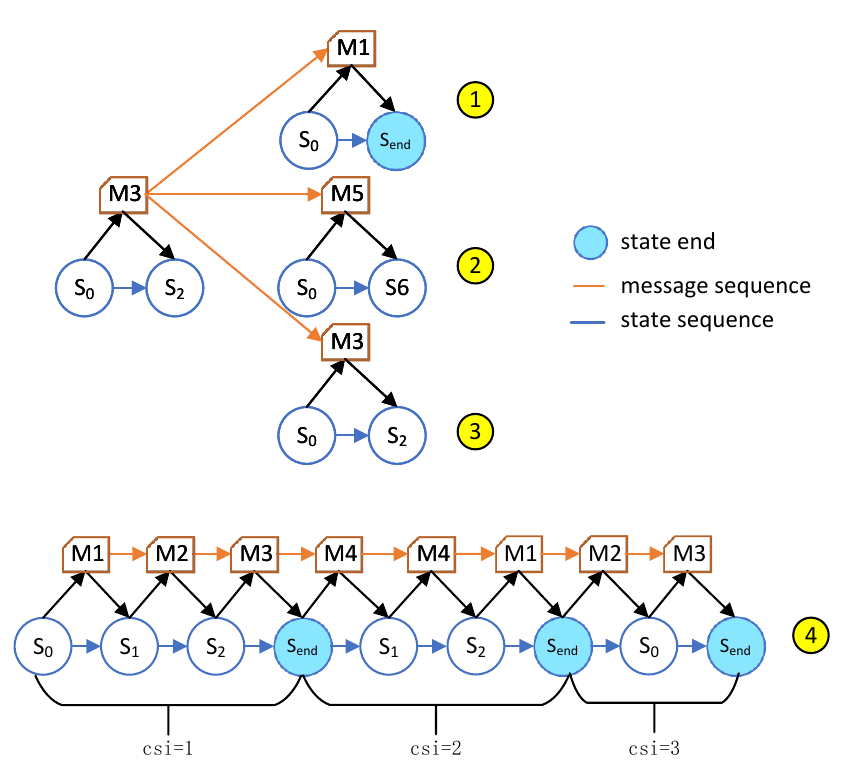}
    \caption{Message Sequence Construct}
    \label{fig:constructmessagesequence}
\end{figure}
In SMGFuzz, algorithm MSC is employed to construct the message sequence.

\begin{algorithm}[h]
    \label{alg:msc}
    \caption{Message Sequence Construct (MSC)}
    \begin{algorithmic}[1]
    \State \textbf{Input:} $S\!M$(state map),$T$(to add list),$Z$(zero point list),
    \State \textbf{Output:} $M\!S$(message sequence)
    \State $M\!S$ $\gets$ NULL
    \State csi$\gets$ CONSTRUCT\_SEQUENCE\_ID(queue\_cur);
    \State $S\!P$(state point) $\gets$ NULL;
    \If{$T$ == NULL}
        \State $S\!P$ $\gets$ GET\_STATEPOINT($T$);
        \State ADD\_TO\_MESSAGE\_SEQUENCE($M\!S$,$S\!P$)
        \If{$Z$ == NULL}
            \State $S\!P$ $\gets$ GET\_STATEPOINT($T$);
            \State  ADD\_TO\_MESSAGE\_SEQUENCE($M\!S$,$S\!P$)
        \Else
            \State $S\!P$ $\gets$GET\_STATEPOINT($Z$);
            \State ADD\_TO\_MESSAGE\_SEQUENCE($M\!S$,$S\!P$)
        \EndIf
    \Else
        \For{$i$ from 0 to LEN($S\!M$) -1}
        \State$S\!P$ $\gets$ GET\_STATEPOINT\_FRO\_MAP($M\!S$, $i$);
            \If{$S\!P$.construct\_sequence\_id == csi}
                \State ADD\_TO\_MESSAGE\_SEQUENCE($M\!S$, $S\!P$);
                \If{$S\!P$.message\_end == TRUE}
                    \State break;
                \EndIf
            \EndIf
        \EndFor
    \EndIf        
    \end{algorithmic}
\end{algorithm}

According to the classification of nodes in SMGFuzz, if there are \texttt{POINT\_TO\_ADD} type nodes in the current queue, it is necessary to prioritize the use of \texttt{POINT\_TO\_ADD} type nodes to construct the message sequence, forming valid state transitions in order to add \texttt{POINT\_TO\_ADD} type nodes to the statemap. As shown in the figure, the  $\langle S_{\text{0}}, S_{\text{2}}\rangle$ node is a \texttt{POINT\_TO\_ADD} type node. At this point, if there are \texttt{POINT\_TO\_ADD} type nodes present, \texttt{POINT\_ZERO} type nodes should be prioritized as subsequent nodes to increase the likelihood of forming valid state transitions. When there are no \texttt{POINT\_ZERO} type nodes present but there are other \texttt{POINT\_TO\_ADD} type nodes, one of the other \texttt{POINT\_TO\_ADD} type nodes should be randomly selected as the subsequent node. If there are no \texttt{POINT\_ZERO} type nodes and no other \texttt{POINT\_TO\_ADD} type nodes present, then the node itself should be repeated to form the message sequence for subsequent testing.

If there are no \texttt{POINT\_TO\_ADD} type nodes in the current queue, the message sequence is constructed based on $S_{\text{dst}}$ and csi (\texttt{construct\_sequence\_id}). The current queue maintains a state\_point\_list, where each state\_point represents a valid state transition triggered by the execution of seeds, and the order is the same as the last fuzzing execution. 
Since SMG defines $S_{\text{end}}$, the state transition sequence is divided into different subsequences by  $S_{\text{end}}$, and each subsequence is represented by a \texttt{sequence\_id}.
 Algorithm MSC traverses the \texttt{state\_point\_list} from the beginning, and when the current node's \texttt{sequence\_id} equals csi, it marks the current node as the initial node of the message sequence and starts constructing the message sequence from this node until encountering a  $S_{\text{end}}$ node. For the same seed, when all nodes in the subsequence corresponding to csi have been mutated, SMG increases csi to start fuzzing the next subsequence.

\subsubsection{Reverse State Selection }
The process of selecting messages for mutation based on statemap in the reverse state selection algorithm can be divided into 4 steps, as shown in Algorithm RSS and Fig.\ref{fig:RSS}:

\begin{figure}[h]
    \centering
    \includegraphics[width=0.48\textwidth]{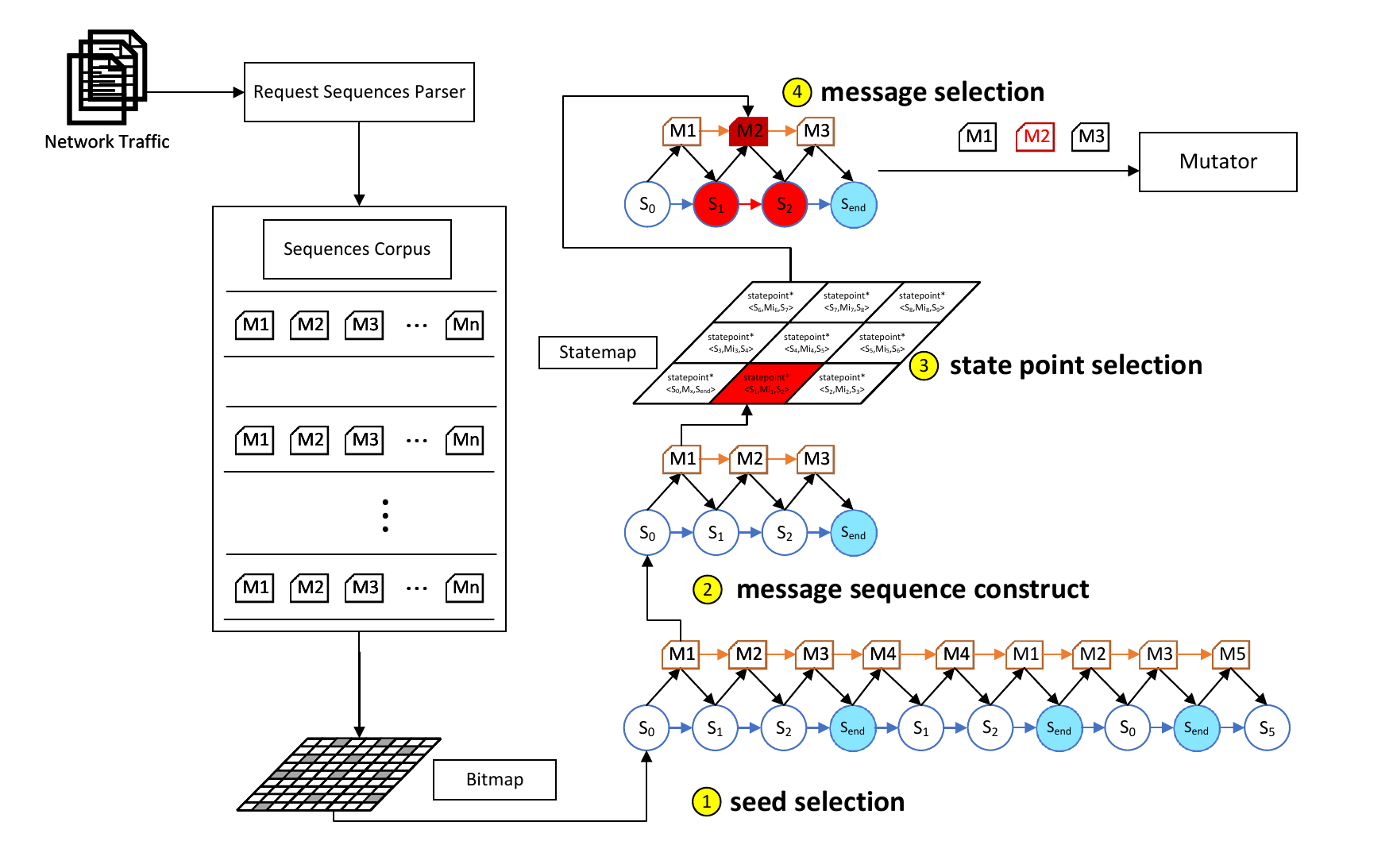}
    \caption{Reverse State Selection (RSS)}
    \label{fig:RSS}
\end{figure}

\textbf{Seed Selection}:Based on the global bitmap information and the bitmap information of seeds, select the seed to be used for this test. SMG relies on the probability model inherited from AFL\cite{afl} to select the optimal seed during the seed selection process.

\textbf{Message Sequence Construction}: Based on the message sequence corresponding to the seed after execution and the type of statepoint obtained, use the algorithm described in Section 3.1 to construct the message sequence required for this test. From the constructed message queue, we can obtain the set of all statepoints S that this test can cover.

\textbf{State Point Selection}: Based on the set of statepoints S corresponding to the test message queue, select the optimal statepoint as the test target according to heuristic rules.

\textbf{Message Selection}: Select the message that can trigger the valid state transition corresponding to the statepoint, mutate it, and complete the subsequent testing.

\begin{algorithm}[!h]
    \label{alg:RSS}
    \caption{Reverse State Selection (RSS)}
    \begin{algorithmic}[1]
    \State \textbf{Input:}$S$(Init Seeds)
    \State \textbf{Output:}$T$(Crashing Inputs)
    \State $T_x$ $\gets$ NULL
    \State $T$ $\gets$ $S$
    \Repeat
        \State $t$ $\gets$ SEED\_SELECTION($T$)
        \State $p$ $\gets$ ASSIGN\_ENERGY($t$)
        \For{$i$ from 1 to $p$}
            \State $M\!S$ $\gets$ MSC($t$)
            \State $S$ $\gets$ REACHABLE\_STATE($M\!S$)
            \State $s$ $\gets$ STATE\_POINT\_SELECTION($S$)
            \State $t^\prime$ $\gets$ MUTATE\_MESSAGE\_SELECTION($s$, $M\!S$)
            \If{$t^\prime$}
                \State add $t^\prime$ to $T_x$
            \ElsIf{ISINTERESTING($t^\prime$)}
                \State add $t^\prime$ to $T$
            \EndIf
        \EndFor
    \Until{timeout reached or abort-signal}
    \end{algorithmic}
\end{algorithm}

\section{EVALUATION}

\subsection{Implementation}
We implemented SMGFuzz based on AFLNet\cite{pham2020aflnet}. Similar to AFLNet\cite{pham2020aflnet}, we use return codes to identify protocol states. The reverse state selection method based on statemap is also applicable to other state recognition methods such as Stateafl\cite{natella2022stateafl} and SGFuzz\cite{ba2022stateful}. We believe that more accurate state recognition methods can similarly improve the performance of the reverse state selection method based on statemap.

\subsection{Experimental setup}
We chose the well-known stateful network protocol fuzzing benchmark ProFuzzBench as our testing target. We compared SMGFuzz with the open-source grey-box protocol fuzzing tools AFLNet\cite{pham2020aflnet} and AFLNwe\cite{aflnwe}.AFLNet is the first grey-box fuzzer based on the SCGF method, which uses the protocol's response messages to distinguish states. AFLNwe is a stateless-guided grey-box fuzzer that modifies AFL's file input mode to network protocol socket transmission mode.

To gain further insight into the impact of SMGFuzz's Message Construct algorithm on test case execution speed, we conducted local experiments targeting the live555 test case. SMGFuzz and AFLNet were run separately for 24 hours, with the execution speed recorded hourly.

We used Snapfuzz to conduct acceleration tests on both SMGFuzz and AFLNet, which seemed to indicate that the algorithm employed by SMGFuzz is similarly effective for this type of fuzzer.

Experiments in \ref{basicblockcoverage} and \ref{heuristicalgorithmimpact} were conducted on a server with 56 cores and 225 GB of RAM, equipped with two Xeon E5-2690 processors and running Ubuntu 20.04. Experiments in \ref{testcaseexecutionspeed} and \ref{snapspeedup} were performed on a separate computer, with i7-12700F processors and 16G RAM. Each experiment was run for 24 hours and repeated 4 times to ensure statistical significance of the results.

\subsection{Research Questions}
To investigate the effectiveness of the Stateful protocol fuzzing with statemap-based reverse state selection method, we propose the following questions:

\textbf{RQ1.}How does the selection of termination states impact the efficiency of fuzz testing?

\textbf{RQ2.}Is the overall fuzz testing efficiency of SMGFuzz superior to aflnet and aflnwe?

\textbf{RQ3.}What is the impact of heuristic algorithms on the final testing effectiveness of SMGFuzz?

\textbf{RQ4.}How does the construction of Message Sequences affect the speed of test case execution?

\textbf{RQ5.}Can the statemap-based reverse state selection algorithm be applied to other SCGF tools?

\begin{figure*}[!h]
    \centering
    \includegraphics[width=\textwidth]{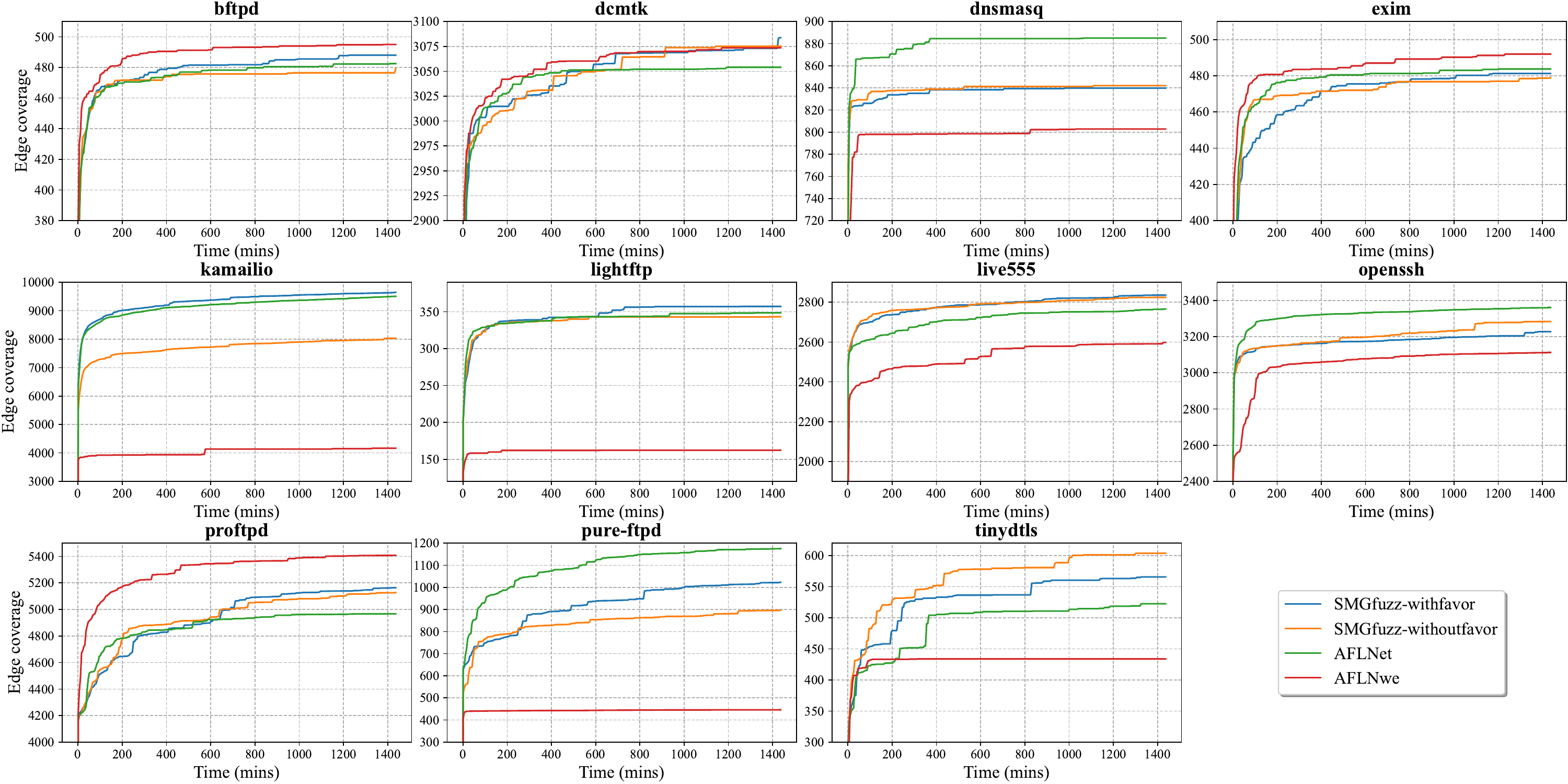}
    \caption{Architecture of SMGFuzz}
    \label{fig:overall}
\end{figure*}

\begin{table*}[!h]
    \label{basicblocks}
	\centering
	\renewcommand{\arraystretch}{1.2}
	\caption{Average Edge Coverage of Each Fuzzer in 24 Hours}
	\begin{tabular}{l|r|r|rrr|rrr}
		\toprule
		
		\multirow{2}{*}{\textbf{Subject}} &
		\multicolumn{1}{c|}{\multirow{2}{*}{\textbf{AFLNet}}} &
		\multirow{2}{*}{\textbf{AFLNwe}} &
		\multicolumn{3}{c|}{\textbf{SMGFuzz  without favor}} &
		\multicolumn{3}{c}{\textbf{SMGFuzz  with favor}} \\\cline{4-9}
		&
		\multicolumn{1}{c|}{} &
		&
		\multicolumn{1}{c}{\textbf{Without favor}} &
		\multicolumn{1}{c}{\textbf{Improv}} &
		\multicolumn{1}{c|}{\textbf{$\bm{\hat{A}_{12}}$}} &
		\multicolumn{1}{c}{\textbf{With favor}} &
		\multicolumn{1}{c}{\textbf{Improv}} &
		\multicolumn{1}{c}{\textbf{$\bm{\hat{A}_{12}}$}} \\\hline \hline
		bftpd     & 484.00  & 495.00  & 479.00  & -1.03\%  & 0.25                  & 488.00  & 0.83\%   & 0.69                  \\
		dcmtk     & 3054.00 & 3067.00 & 3075.25 & 0.70\%   & 0.88                  & 3083.75 & 0.97\%   & 0.81                  \\
		dnsmasq   & 885.00  & 802.75  & 842.00  & -4.86\%  & 0.13                  & 839.75  & -5.11\%  & 0.19                  \\
		exim      & 483.75  & 490.25  & 479.00  & -0.98\%  & 0.38                  & 481.25  & -0.52\%  & 0.38                  \\
		kamailio  & 6592.50 & 4163.25 & 9692.00 & 47.02\%  & 0.94                  & 9650.00 & 46.38\%  & 1.00                  \\
		lightftp  & 349.50  & 162.25  & 343.25  & -1.79\%  & 0.06                  & 357.50  & 2.29\%   & 0.63                  \\
		live555   & 2766.00 & 2598.00 & 2824.50 & 2.11\%   & 1.00                  & 2836.25 & 2.54\%   & 0.94                  \\
		openssh   & 3361.25 & 3112.75 & 3283.25 & -2.32\%  & 0.06                  & 3227.00 & -3.99\%  & 0.00                  \\
		proftpd   & 4967.25 & 5407.00 & 5127.00 & 3.22\%   & 0.81                  & 5164.00 & 3.96\%   & 0.88                  \\
		pure-ftpd & 1174.75 & 446.25  & 895.50  & -23.77\% & 0.25                  & 1023.00 & -12.92\% & 0.06                  \\
		tinydtls  & 522.50  & 433.75  & 603.50  & 15.50\%  & 1.00                  & 565.50  & 8.23\%   & 0.69                  \\\hline \hline 
		AVG       & 2240.05 & 1925.30 & 2513.11 & 12.19\%  & - & 2519.64 & 12.48\%  & - \\
		\bottomrule
	\end{tabular}
\end{table*}

\subsection{End State Selection (RQ1)}
\label{EndStateSelection}
It might be helpful to note that SMGFuzz requires the use of end states to construct the message sequence and determine the state point type. It would be advisable, therefore, at the beginning of fuzz testing, to pass the end state as an initial parameter to SMGFuzz. In order to ensure fairness in comparison with AFLNet, we used AFLNet to run the initial seeds from ProFuzzBench. Following the construction of the state diagram by AFLNet for the initial seeds, we proceeded to select the end states from the initial state diagram as the input for SMGFuzz to conduct the tests.

\subsection{Edge Coverage Evaluation (RQ2)}

\label{basicblockcoverage}
In Fig. \ref{fig:overall} and Table \ref{basicblocks}, we present the average edge coverage of AFLNet, AFLNwe, and SMGFuzz after 24 hours of fuzz testing, both with and without favour. We hope that this information will be of use to you. During fuzz testing of dcmtk, kamailio, live555, proftpd, and tinydtls, SMGFuzz demonstrated a slightly higher edge coverage than AFLNet. By incorporating the heuristic method (SMGFuzz with favour), SMGFuzz showed promising results in edge coverage for bftpd and lightftpd, and also demonstrated a notable improvement in pure-ftpd. Overall, these results suggest that SMGFuzz with favour may offer a competitive edge over SMGFuzz without favour in edge coverage.

It is worth mentioning that in the fuzz testing of bftpd and proftpd, AFLNwe demonstrated superior edge coverage compared to the stateful AFLNet. It might be observed that AFLNwe, being a stateless greybox fuzzing tool based on bitmap, retains bitmap guidance for seed selection. Similarly, SMGFuzz, which also considers bitmap guidance for seed selection, achieved comparable edge coverage to AFLNet in fuzz testing of bftpd and proftpd. In the fuzz testing of dnsmasq and pure-ftpd, AFLNet achieved better basic block coverage. However, SMGFuzz, which considers the state of the SUT, showed promising results that could be further improved.

In conclusion, it can be said that SMGFuzz has the potential to enhance overall edge coverage by leveraging the mapping relationship between bitmap and statemap, thus making better use of both bitmap and statemap information.

In Table \ref{uniquecrashes}, we present the total number of crashes discovered after 24 hours of fuzz testing by AFLNet\cite{pham2020aflnet}, AFLNwe\cite{aflnwe}, and SMGFuzz. Among them, SMGFuzz discovered the highest number of unique crashes in the fuzz testing of live555, kamailio, and dnsmasq. Compared to AFLNet\cite{pham2020aflnet}, SMGFuzz showed a 50\% increase in the total number of unique crashes discovered.
\begin{table}[h]
    \centering
    \caption{Total Number Of Unique Crashes Found By Each Fuzzer in 24 Hours}
    \label{uniquecrashes}
    \begin{tabular}{cccc}
    \hline
    program  & AFLNet & AFLNwe & SMGFuzz \\ \hline
    exim   & 0      & 0      & 0       \\
    dcmtk    & 22     & 0      & 20      \\
    live555  & 480    & 617    & 815     \\
    kamailio & 1      & 0      & 16      \\
    proftpd  & 0      & 0      & 0       \\
    dnsmasq  & 209    & 175    & 218     \\ \hline
    Average  & 712    & 792($11.24\%\uparrow$ )    & 1069($50.14\%\uparrow$ )    \\ \hline
    \end{tabular}
\end{table}

\subsection{Heuristic Algorithm Impact (RQ3)}
\label{heuristicalgorithmimpact}
In the favourable state, AFLNet conducts five rounds of testing on each state, with the aim of accumulating sufficient data before starting to optimise the state selection. It selects states that are more likely to result in the generation of new branches for fuzz testing. SMGFuzz places a high value on ensuring comprehensive coverage of the bitmap seed set. Once a round of bitmap coverage has been completed, SMGFuzz employs a similar heuristic function to that used by AFLNet in order to select the most appropriate state from those covered by the seeds for fuzz testing.

As can be seen in Table 1, the heuristic algorithm has the potential to enhance SMGFuzz's edge coverage for bftpd, lightftp, and pure-ftpd. Furthermore, some improvement was observed in the testing of DCMTK and EXIM, which contributed to enhancing SMGFuzz's overall edge coverage.

\subsection{Testcase execution speed (RQ4)}
\label{testcaseexecutionspeed}

\begin{figure}[h]
    \centering
    \includegraphics[width=0.48\textwidth]{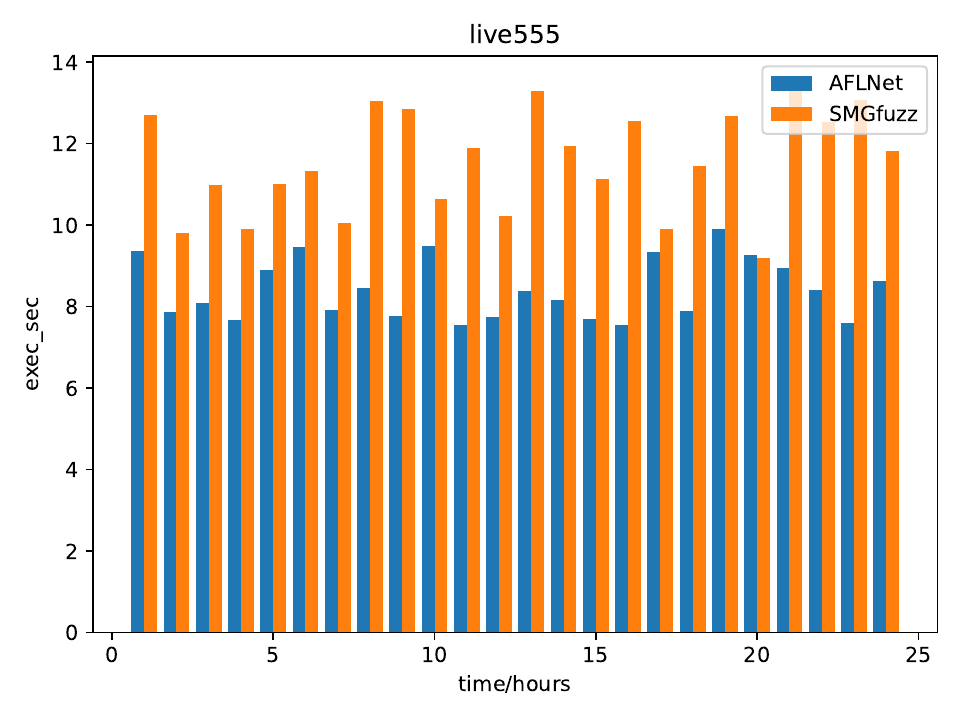}
    \caption{Testcase execution speed in 24 hours}
    \label{fig:live555_24h }
\end{figure}

In Fig.\ref{fig:live555_24h }, we illustrate the comparison of test case execution speeds between AFLNet\cite{pham2020aflnet} and SMGFuzz over 24 hours. Due to SMGFuzz's design of different state types and optimization of message sequence construction methods, it has improved the efficiency of executing individual test cases. During the 24-hour fuzz testing of live555, SMGFuzz achieved an average test case execution speed of 11.32 execs/sec, while AFLNet's average test case execution speed was 8.08 execs/sec. SMGFuzz achieved a 40.2\% improvement in test case execution speed.

\subsection{Snap Speedup(RQ5)}
\label{snapspeedup}

SnapFuzz\cite{andronidis2022snapfuzz}is a fuzz testing tool based on AFLNet. It provides a robust architecture that converts slow asynchronous network communication into fast synchronous communication using UNIX domain sockets. It accelerates all file operations by redirecting file operations to an in-memory file system and eliminates the need for many fragile modifications, such as configuring time delays or writing cleanup scripts, along with other improvements.

From an implementation perspective, SnapFuzz retains all of AFLNet's state selection algorithms, with the exception of the network interface implementation, which has been modified. As a result, we decided to replace SnapFuzz's original state selection algorithm with SMGFuzz and compare it with the original algorithm. We conducted a 24-hour fuzz testing session using SnapFuzz in long mode, with live555 selected as the test target.

The experimental results suggest that SnapFuzz, when using the SMGFuzz algorithm, achieved an average of 401.08 test cases per second, which represents an improvement over the 354.70 test cases per second achieved by the original algorithm. Furthermore, we are pleased to report that the total number of paths discovered in 24 hours was 2183, which represents a 6.7\% increase compared to the 2046 paths found with the original algorithm. It may be of interest to note that the SMGFuzz algorithm has also been shown to be effective when applied to other tools based on AFLNet, with similar performance improvements.

\section{RELATED WORK}
According to the stateful grey-box methods used in protocol fuzzers, grey-box protocol fuzzers can be classified into five categories: based on protocol message feedback, based on protocol interaction message analysis, based on SUT  code analysis, based on SUT memory dynamic detection, and based on protocol description document grammar learning. Table 1 summarizes the main features of representative tools in each category. In the following sections, we will analyze in detail the technical characteristics of tools in each category.

\begin{table*}[!h]
\centering
\caption{Technics used in stateful grey-box fuzzing}
\label{relatedwork}
\begin{tabular}{cccc}
\hline
Technics & Fuzzer & Fundamental tools & Scope \\
\hline
Response Message & AFLNet\cite{pham2020aflnet} & afl & Grey-box fuzzing\\
SUT Code Static Analysis& SGFuzz\cite{ba2022stateful} & libfuzzer\cite{serebryany2016continuous} & Grey-box fuzzing \\
\multirow{2}{*}{Protocol Interaction Message Analysis} & BLEEM\cite{luo2023bleem} & --& Black-box fuzzing \\
& Pulsar\cite{gascon2015pulsar} & PRISMA\cite{krueger2012learning} & Black-box fuzzing \\
SUT Memory Dynamic Detection & Stateafl\cite{natella2022stateafl} & afl & Grey-box fuzzing \\
\hline

\end{tabular}

\begin{tablenotes}
    \footnotesize
    \item Note: The technical method refers to the technical means used by the tool to construct the state machine of the System Under Test (SUT); Basic tools refer to the basic tools used in the implementation of the tool; Test case generation method refers to the method used by the tool to generate test cases; Scope refers to the scope of the fuzzing techniques adopted by the tool.
\end{tablenotes}
\end{table*}

\subsection{Response Message}
The state guidance method based on protocol message return information continuously captures the response messages of SUT, constantly updating the observed protocol state machine to guide the protocol fuzzing process. Taking AFLNet\cite{pham2020aflnet} as an example, AFLNet\cite{pham2020aflnet} reads the server's response messages into specific buffers, extracts the protocol's status response codes, and continuously updates its maintained protocol state machine model based on these response codes. If there are new states in the server response codes, AFLNet\cite{pham2020aflnet} adds these new states to its state machine. AFLNet\cite{pham2020aflnet} selects the target states for the fuzzer from the protocol state machine model using heuristic methods. AFLNet\cite{pham2020aflnet} prioritizes states based on the number of message sequences generated by the executed states, where states with more generated message sequences have lower priorities, allowing the fuzzer to focus on less frequently executed protocol states.

The state guidance method based on protocol message return information relies on return information for analysis, making it unsuitable for protocols without return messages, such as HTTP/2 protocol. Consequently, its applicability is limited, restricting its usage in certain contexts. However, as the earliest proposed state guidance method, it laid the foundation for subsequent research and developments in this field.

\subsection{Protocol Interaction Message Analysis}
The state-guided method based on protocol interaction message analysis typically captures protocol interaction messages within a certain time window, analyzes the captured interaction data using specific algorithms, thereby constructing a protocol state machine, and further guiding the fuzzing testing process.

Pulsar \cite{gascon2015pulsar} first captures all traffic between the client and server, identifies sessions from it, and intervenes manually in the communication process between the client and server to ensure that the traffic contains all the functionalities of the protocol as much as possible. Then, it clusters the identified protocols and constructs different clustering models for text protocols and binary protocols, initially obtaining the internal format of the protocol. Pulsar \cite{gascon2015pulsar} tracks the entire interaction process of the protocol and, based on the probability of message occurrences in the protocol, constructs it into a second-order Markov model, further minimized to a deterministic finite automaton (DFA). At this point, by establishing the mapping from the internal format of the protocol to the state machine, Pulsar\cite{gascon2015pulsar} has obtained a complete description of the protocol. Pulsar further defines a fuzzing subgraph (FS) to implement state guidance. It assigns a consumable mask to each message template in the FS. When selecting a state for fuzzing testing, the consumed mask of the used message template decreases accordingly, and the weight of this state also decreases accordingly, allowing fuzzing testing to continue for other states. This state selection algorithm ensures that all states are tested with the same probability, thereby avoiding the long-term occupation of testing resources by certain states.

Unlike Pulsar\cite{gascon2015pulsar} , BLEEM\cite{luo2023bleem} needs to act as a communication proxy between the client and server, intercepting and collecting communication data between them. BLEEM\cite{luo2023bleem}  uses Scapy to parse packets, recognize packet formats, and then extract message semantics to construct abstract packet sequences, which are then converted into state tracking and further merged into a System State Tracking Graph (SSTG). During the testing process, BLEEM\cite{luo2023bleem}  dynamically maintains the SSTG by intercepting normal communication packets sent between the client and server, obtaining a more precise description of the protocol state machine. Based on the accurate SSTG, BLEEM\cite{luo2023bleem}  designs a heuristic state guidance algorithm that selects available packets, constructs input sequences, and gradually achieves a comprehensive traversal of the SSTG from high-density state areas to low-density state areas.

The state-guided method based on protocol interaction message analysis avoids the limitation of relying solely on single return information for analysis. It can introduce advanced analysis algorithms to obtain accurate protocol state machine models, thereby further guiding the generation of fuzzy test cases accurately and achieving good testing results.

\subsection{SUT Code Static Analysis}
The state-guided method based on static analysis of SUT code primarily involves static analysis of the SUT during its compilation process to extract the state machine model of the target protocol, thereby further guiding fuzz testing. A typical representative of this method is SGFuzz\cite{ba2022stateful}, which identifies state variables and switch or if statements in the target code to analyze the state machine of the target protocol. Based on this analysis, SGFuzz\cite{ba2022stateful} constructs the State Transition Tree (STT) of the target protocol, where each node of the STT represents the value of state variables during program execution, and edges represent state transitions. To effectively explore the protocol state space in guiding fuzz testing, SGFuzz\cite{ba2022stateful} designs corresponding guidance algorithms based on the STT. First, it assigns a "hit count" to each node of the STT, recording the number of inputs traversing that node. The algorithm assumes that neighbors of nodes with low hit counts have more states to explore, thereby increasing the exploration probability of nodes with low hit rates. Secondly, it assigns higher priority to effective state transitions because state transitions corresponding to expected protocol behavior typically mutate easily into other ineffective state transitions, representing protocol error handling. Through these two aspects, SGFuzz\cite{ba2022stateful} can fully utilize the STT to guide fuzz testing for complete coverage of protocol states.

The state-guided method based on static analysis of SUT code relies on the source code of the SUT, making it difficult to extract state machines for commercial protocol software, thus limiting its applicability.

\subsection{SUT Memory Dynamic Detection}
The state-guided method based on dynamic memory detection of SUT primarily involves monitoring the memory and function calls of the SUT and analyzing the behavior of sending and receiving messages to dynamically construct and continuously update the protocol state machine model, thereby guiding the fuzz testing process. Taking Stateafl\cite{natella2022stateafl} as an example, Stateafl\cite{natella2022stateafl} compiles detection probes into relevant code during SUT compilation to track memory allocation, deallocation, message sending and receiving, and service startup. Through a dynamic state machine construction algorithm, Stateafl\cite{natella2022stateafl} continuously updates the protocol state machine of the SUT. Based on this, Stateafl\cite{natella2022stateafl} selects target states according to the number of mutations, the number of selections, and the increase in path coverage in the state machine. The probability of selecting the target state is inversely proportional to the number of mutations and selections, and directly proportional to the increase in path coverage.

The state-guided method based on dynamic memory detection of SUT avoids potential errors introduced by extracting protocol state variables in static analysis methods and can obtain a more accurate protocol state machine model. However, in some specific scenarios (e.g., fuzz testing of IoT device protocols), it may be challenging to achieve dynamic analysis of the SUT, making it impossible to obtain the protocol state machine model through dynamic analysis.

\section{CONCLUSION}
In this article, we summarized the shortcomings of existing SCGF methods in representing protocol state machines and constructing message sequences. To enhance the efficiency of SCGF, we proposed SMGFuzz, which utilizes the reverse state selection method and designed Statemap for discretizing the representation of protocol state machines. Statemap simplifies the method of representing state machines, thereby improving the efficiency of protocol state analysis. Furthermore, we optimized the message sequence construction method in SCGF by designing different types of state points, thereby reducing the length of message sequences and further improving the efficiency of test case execution. In the 24-hour fuzz testing experiments, SMGFuzz achieved a 12.48\% increase in basic block coverage, a 50.1\% increase in the total number of unique crashes discovered, and a 40.2\% increase in test case execution speed.







\bibliographystyle{elsarticle-num}  
\bibliography{elsarticle-template-num}

\end{document}